\documentclass{article}

\usepackage{arxiv}

\usepackage[utf8]{inputenc} 
\usepackage[T1]{fontenc}    
\usepackage{hyperref}       
\usepackage{url}            
\usepackage{booktabs}       
\usepackage{amsfonts}       
\usepackage{nicefrac}       
\usepackage{microtype}      
\usepackage{lipsum}		
\usepackage{graphicx}
\usepackage{natbib}
\usepackage{tcolorbox}
\usepackage{listings}
\usepackage{doi}

\title{Co-Designing a Chatbot for Culturally Competent Clinical Communication: Experience and Reflections}

\author{ \href{https://orcid.org/0000-0002-9975-8844}{\includegraphics[scale=0.06]{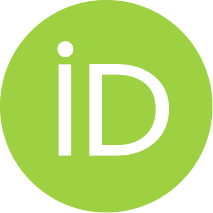}\hspace{1mm}Sandro Radovanović}\\
	Faculty of Organizational Sciences\\
	University of Belgrade\\
	Belgrade, Serbia \\
	\texttt{sandro.radovanovic@fon.bg.ac.rs} \\
	\And
	\href{https://orcid.org/0000-0002-8651-1148}{\includegraphics[scale=0.06]{orcid.pdf}\hspace{1mm}Shuangyu Li} \\
	Faculty of Arts and Humanities\\
	King's College London\\
	London, United Kingdom \\
	\texttt{shuangyu.li@kcl.ac.uk} \\
}

\renewcommand{\shorttitle}{Co-Designing a Chatbot for Culturally Competent Clinical Communication}

\hypersetup{
pdftitle={Co-Designing a Chatbot for Culturally Competent Clinical Communication Experience and Reflections},
pdfauthor={Sandro Radovanović, Shuangyu Li},
}

\begin{document}
\maketitle

\begin{tcolorbox}[colback=yellow!10!white,colframe=red!75!black,title=Research in Progress, sharp corners=south]
This manuscript reports preliminary findings from an ongoing research project. The current version reflects initial testing with a limited sample size and early-stage chatbot interactions. Further user engagement and iteration are required to generate results that can support more robust and generalisable claims.
\end{tcolorbox}
\bigskip

\begin{abstract}
Clinical communication skills are essential for preparing healthcare professionals to provide equitable care across cultures. However, traditional training with simulated patients can be resource intensive and difficult to scale, especially in under-resourced settings. In this project, we explore the use of an AI-driven chatbot to support culturally competent communication training for medical students. The chatbot was designed to simulate realistic patient conversations and provide structured feedback based on the ACT Cultural Competence model. We piloted the chatbot with a small group of third-year medical students at a UK medical school in 2024. Although we did not follow a formal experimental design, our experience suggests that the chatbot offered useful opportunities for students to reflect on their communication, particularly around empathy and interpersonal understanding. More challenging areas included addressing systemic issues and historical context. Although this early version of the chatbot helped surface some interesting patterns, limitations were also clear, such as the absence of nonverbal cues and the tendency for virtual patients to be overly agreeable. In general, this reflection highlights both the potential and the current limitations of AI tools in communication training. More work is needed to better understand their impact and improve the learning experience.
\end{abstract}


\section{Introduction}
\label{sec:introduction}

Effective clinical communication is a core skill for healthcare professionals, yet many students continue to find it challenging to master, even after years of training \citep{ref10}. One potential cause mentioned in the literature is inadequate communication, as it can impact patient safety and satisfaction and contribute to moral distress and burnout among healthcare professionals \cite{ref17}. As awareness of the role cultural diversity plays in healthcare care increases \citep{ref16}, medical schools have been working to integrate cultural competence into their curricula \citep{ref15}. However, communication breakdowns in culturally and linguistically diverse settings remain common, and the learning opportunities to address these challenges can be limited. Given these limitations, there is a pressing need for scalable, cost-effective and adaptive training solutions to better prepare future physicians for culturally competent care \citep{ref13}.

Our experience suggests that traditional training methods, such as using standardized patients, offer valuable experiential learning, but are often resource-heavy and not easily scalable. These methods depend on trained actors, significant scheduling effort, and substantial financial support. In our setting, these constraints meant that students had relatively few opportunities to participate in a broad range of cultural scenarios. In addition, we found that the feedback from standardized patients could be inconsistent, making it harder for students to reflect on and systematically improve their skills.

Motivated by these observations, we explore whether a digital tool could complement existing training by offering additional practice opportunities in a more accessible and flexible way. Specifically, we co-developed a beta version of a chatbot using OpenAI’s GPT-4o model. The chatbot aimed to simulate culturally nuanced clinical consultations and provide structured feedback based on the ACT Cultural Competence model, published as a BEME Guide \citep{ref12}. However, rather than trying to replicate the entire experience of human interaction, our goal was to design a low-pressure, text-based tool that students could use to practice communicating with diverse patient profiles. The tool was developed collaboratively by a medical educator, a machine learning specialist, and medical students, drawing on our collective insights to guide its structure and content.

This paper shares our reflections from developing and piloting the chatbot within a clinical communication module at a large UK medical school. We do not present this as a definitive solution, but rather as an exploratory step toward using AI tools to complement and expand cultural communication training. Our aim is to contribute to ongoing conversations about how such tools might be meaningfully integrated into medical education, particularly in ways that support both reflection and practice.

Key features of the chatbot include:
\begin{itemize}
    \item Dynamic patient responses that challenge students to adapt based on cultural and contextual cues;
    \item Feedback organized around the ACT framework, offering reflections on cultural awareness, critical thinking, and equitable care delivery;
    \item Actionable suggestions to help students gradually improve their communication behaviours; and
    \item The ability for students to practice at any time, without the constraints of scheduling or limited access to faculty support.
\end{itemize}

Although this tool showed early signs of promise, we also encountered several limitations. The chatbot’s reliance on written text means it lacks non-verbal cues like tone or facial expressions, which are central to real-world communication. However, we noticed that the slower and more deliberate pace of text-based interaction gave some students space to reflect on their responses, something that may be particularly helpful for those early in their training. This format also resembles some interactions on digital health platforms such as Babylon Health or Ada Health, where clinicians often communicate asynchronously in writing.

The remainder of the paper is structured as follows. Section \ref{sec:background} provides background on how the ACT Culturally Competent Model was transformed into a Culturally Competent Communication Guide for the chatbot. Section \ref{sec:chatbot_design} gives a technical description of the chat bot, while Section \ref{sec:results} provides the initial results and analysis. The main idea is not solely to show how the chatbot improved the communication but to show what can be done and presented back to the student. Finally, we conclude the paper in Section 5.

\section{Background and Learning Framework}
\label{sec:background}

\begin{figure}[h!]
    \centering
    \includegraphics[width=1\linewidth]{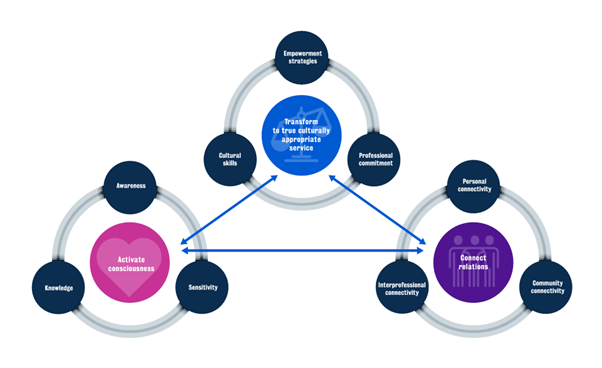}
    \caption{ACT Cultural Competence Model diagram is recreated based on the \citep{ref12}}
    \label{fig:act_model}
\end{figure}

To support students in developing culturally competent communication skills, we grounded our chatbot design in the ACT Cultural Competence Model \citep{ref12} as presented in Figure \ref{fig:act_model}. This model, originally developed for medical education, was adapted into a practical Culturally Competent Communication Guide (CCCG) through collaboration between a medical educator, a machine learning expert, and medical students. The CCCG is used within the chatbot to guide feedback and structure conversations. It is organized around three domains of learning: \textbf{Activating Awareness}, \textbf{Connecting Relations}, and \textbf{Transforming into Culturally Appropriate Care}.

Each domain highlights specific expectations for students, encouraging them to reflect on their practice, engage meaningfully with patients, and develop strategies to support equitable care.

\subsection{Activating Awareness}
\label{subsec:activating_awareness}

In this first domain, students are expected to develop awareness on three levels, namely \textit{intrapersonal}, \textit{interpersonal}, and \textit{systemic}, as well as acquire relevant cultural and critical knowledge.

\textit{Intrapersonal awareness} involves reflecting on your own cultural identity, personal beliefs, and unconscious biases \citep{ref5,ref18}. They are encouraged to ask questions like: ``Am I making assumptions based on my own background?'' or ``How might my worldview differ from the patient's?'' This kind of reflection helps identify how your beliefs could unintentionally influence your clinical decisions or communication style.

\textit{Interpersonal awareness} means recognizing that others can express themselves or interpret clinical situations differently, depending on their cultural context. Effective communication requires openness and curiosity rather than assumptions \citep{ref14}.

\textit{Systemic awareness} asks students to think beyond the individual encounter. They are invited to consider how institutional or social structures could impact patient access, trust, or care-seeking behavior, especially for marginalized groups.

To deepen this awareness, students are encouraged to engage in both \textit{cultural knowledge} (understanding norms, values, and traditions) and \textit{critical knowledge} (questioning how dominant narratives shape healthcare systems). Developing this understanding supports greater cultural sensitivity and empathy, forming the foundation of equitable  patient-centered care.

\subsection{Connecting Relations}
\label{subsec:connecting_relations}

This second domain focuses on building meaningful relationships across three levels: \textit{personal}, \textit{community}, and \textit{interprofessional}.

\textit{Personal connectivity} involves taking a person-centered approach, engaging not only with the clinical issue, but also with the patient's background, family, and lived experience. They are expected to ask questions that show genuine interest and support shared decision making \citep{ref12,ref9}.

\textit{Community connectivity} involves recognizing that health is shaped by the broader social environment. Although students are not expected to solve structural issues, they should be aware of how culture, language, and access to resources can affect care. Recognizing these factors shows a deeper understanding of patients' realities \citep{ref2}.

\textit{Interprofessional connectivity} refers to respectful collaboration with colleagues. Students should show openness to team-based care, including valuing the insights of social workers, interpreters, and community health workers. This also involves clear and inclusive communication within a multidisciplinary setting.

\subsection{Transforming into Culturally Appropriate Care}
\label{subsec:transforming_cac}

This final domain focuses on turning reflection into action. Includes empowerment, cultural communication skills, and professional commitment.

\textit{Empowerment} involves recognizing the power dynamics that exist in interactions. Students are expected to create space for the patient's voice and agency. They might demonstrate this by inviting the patient to share what matters most to them or by explicitly acknowledging the systemic barriers they face. \citep{ref12,ref14}

\textit{Cultural skills} include being able to adapt communication styles, use interpreters effectively, and apply cultural formulation, that is, understanding how culture shapes symptoms, concerns, and care expectations.

\textit{Professional commitment} refers to the willingness to engage in lifelong learning and continuous self-reflection. Students are encouraged to identify areas where they need to grow, seek feedback, and remain open to changing their approach over time.

\subsection{Culturally Competent Communication Guide}
\label{subsec:ccc_guide}

The chatbot was designed not only to simulate clinical interactions, but also to evaluate and guide student responses using structured criteria rooted in the ACT framework. These criteria were translated into conversational logic and feedback mechanisms within the chatbot, allowing it to assess how students engage with key aspects of cultural competence and to provide targeted prompts for self-reflection and growth. This structure informed the chatbot’s logic and evaluation processes and was developed collaboratively by a medical education specialist, a machine learning and AI expert, and medical students.

The development of the Cultural Competence Conversational Guide (CCCG) was a deliberately iterative and interdisciplinary process. Early prototypes of the CCCG chatbot balanced between concise answers and the need for detailed, context-sensitive answers, mimicking the expected knowledge of an actual patient. This trade-off between methodological rigor and usability was navigated through repeated cycles of testing and refinement. The medical educator and machine learning expert worked in tandem, alternately generating, testing, and critiquing chatbot interactions. The educator focused on ensuring that the chatbot remained pedagogically sound and clinically relevant, while the AI expert ensured that the chatbot was implementable and scalable. Each iteration addressed errors in logical inference, misinterpretation, or missed opportunities for feedback. Only when both experts were satisfied with the chatbot’s ability to generate accurate, contextually relevant feedback, the tool deployed for student use.

The adapted framework guides the chatbot’s feedback across three key domains:
\begin{itemize}
    \item \textbf{Activate Consciousness}: This domain encourages students to cultivate self-awareness and recognize the influence of their own cultural identities, assumptions, and biases. It also encourages students to critically reflect on the cultural assumptions embedded in clinical situations and to acknowledge the historical and systemic factors that shape healthcare disparities.
    \item \textbf{Connect Relations}: This domain focuses on building respectful and empathetic communication between cultural differences. Students are expected to demonstrate active listening, cultural humility, and an openness to learning from others. It also assesses how well students create inclusive dialogue and recognize the lived experiences of patients and colleagues.
    \item \textbf{Transform to True Cultural Care}: This domain emphasizes action. It challenges students to propose practical, inclusive solutions that reflect an understanding of cultural dynamics and promote equity in healthcare delivery. Students are encouraged to consider how culturally competent care can be embedded in practice and supported at a systemic level.
\end{itemize}

Each domain contains specific behaviors and indicators for which the chatbot looks, such as statements of bias, critical reflection, historical contextualization, inclusive language use, empathy, and equity-driven thinking. These evaluation criteria serve as a practical bridge between the ACT Cultural Competence model and real-world communication behaviors, operationalized through the CCCG.

The chatbot leverages the evaluative power of AI not to replace human instruction, but to augment it. We aim to provide immediate and structured feedback based on CCCG. Therefore, the chatbot can help students reflect on the entire communication more efficiently and achieve learning objectives faster. Combining AI and BERT-based models, one can detect subtle changes in language and emotions, as well as review prior responses, thus helping to form a formative assessment of communication. Rather than waiting for human instructors, students can engage in self-paced learning supported by just-in-time insights into their cognitive and communicative performance.

For a detailed description of these criteria and their implementation within the chatbot, we refer to Appendix \ref{app:cc_act}.

\section{ChatBot Design}
\label{sec:chatbot_design}

\begin{figure}[h!]
    \centering
    \includegraphics[width=1\linewidth]{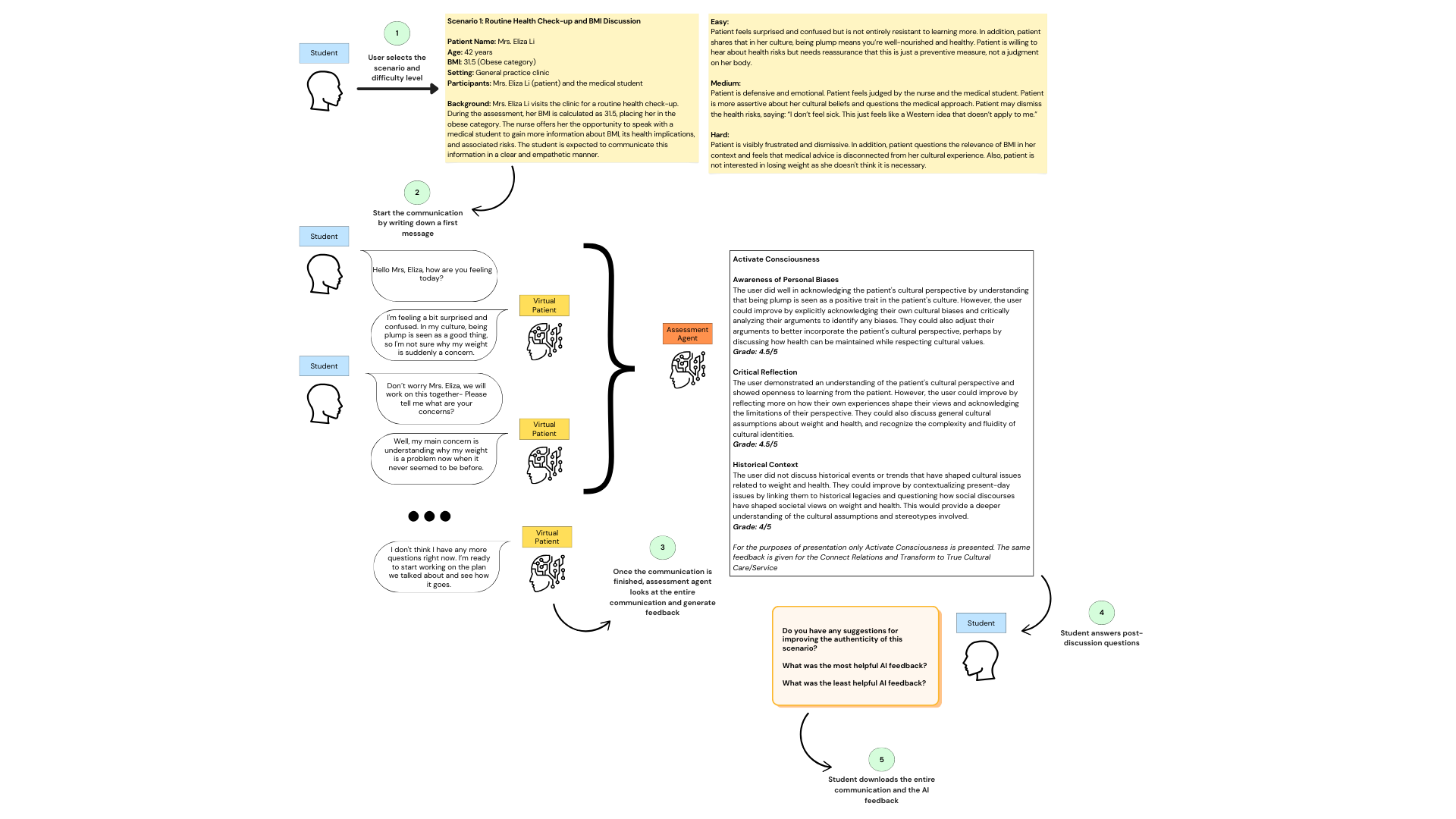}
    \caption{Chatbot experimental flow design }
    \label{fig:chatbot_flow}
\end{figure}

In this study, we developed a Streamlit-based chatbot application to support the development of clinical communication skills, particularly in culturally complex scenarios. The application was piloted with medical students during the 2024/2025 academic year. A full example of chatbot communication is presented in Figure \ref{fig:chatbot_flow}.

\textbf{Chatbot Communication}. The students were first introduced to the ACT framework and instructed to test their communication skills using the chatbot. The learning process involved the following steps:

\begin{itemize}
  \item The students selected a clinical communication scenario to explore, as well as the complexity level. If unsure, they were encouraged to choose one and read its description. Each scenario was designed to reflect realistic and dynamic challenges drawn from multiple years of experience in clinical communication teaching. The scenarios included socially awkward moments, culturally sensitive topics, and complex patient backgrounds. They were co-developed through a collaborative iterative process that involved a medical education specialist, an AI expert, and students. Each scenario and difficulty level were tested against the CCCG framework to ensure that it targeted relevant elements and refined until all three development groups were satisfied.
  \item After reviewing the scenario, students engaged in a live text-based dialogue with the chatbot that simulates a patient. Each interaction was designed to take 6 to 10 minutes. The students remained anonymous and were allowed up to two interactions per scenario.
\end{itemize}

\textbf{Scenarios}. Each student was anonymized and allowed a maximum of two interactions per scenario. A detailed description of the scenarios is provided in the Appendix \ref{app:scenarios}. The scenarios included:

\begin{itemize}
    \item Routine health check-up and BMI discussion – Communicating with a patient about obesity in a sensitive and empathetic way.
    \item Urinary tract infection in a transgender patient – Ensuring respectful and inclusive communication with a transgender individual while discussing sensitive health issues.
    \item Language barrier in a new patient consultation – Navigating a consultation with a patient who has limited English proficiency and lacks an interpreter.
    \item Chronic pain management and cultural considerations – Respecting cultural beliefs while discussing pain management options with a patient who prefers natural remedies.
\end{itemize}

To simulate realistic patient behavior, each scenario was also tagged with a difficulty level, controlled by assigning distinct emotional and personality traits to the virtual patient.
\begin{itemize}
    \item Easy: Traits like surprise, shyness, confusion, and mild skepticism.
    \item Medium: Traits such as defensiveness, emotional responses, nervousness, and moderate distress.
    \item Hard: Involving strong resistance, disengagement, and deep discomfort or worry.
\end{itemize}

\textbf{Feedback and Assessment}. Upon completion of the communication exercise, the students received structured feedback based on the three elements of the CCCG explained in the previous section. Namely, \textit{Activate Consciousness}, \textit{Connect Relations}, and \textit{Transform to True Cultural Care/Service}.

Each domain was scored using a scale from 1 to 5, where 1 represents a limited demonstration of competence and 5 reflects a highly developed response. The scoring was powered by an LLM-based Assessment Agent, with prompts and evaluation logic carefully tuned in collaboration with a medical educator to ensure the pedagogical validity. The scores were accompanied by qualitative feedback that highlighted strong points, areas for improvement, and actionable suggestions for future performance.

For each of these elements, students received a structured response detailing their communication performance within the selected scenario. Feedback included a grade, as well as specific feedback on the good points of the discussion, what was flawed in the discussion, and suggestions for improvement for future discussions. Instructions on the assessment methodology are provided in the Appendix \ref{app:cc_act}.

\textbf{Chatbot Architecture}. To support the interaction and feedback process, we developed two complementary agents using the OpenAI GPT-4o \citep{ref11} model:

\begin{itemize}
  \item \textbf{Virtual Patient:} This agent adopted the role of a simulated patient, engaging the student with contextually appropriate responses. It was configured with a temperature of 0.7 to promote variability and natural conversation flow. The token output was limited to 500 for the sake of conciseness.

  \item \textbf{Assessment Agent:} Triggered when a student ended the conversation (via a "Stop Communication" button), this agent evaluated the full interaction using structured prompts aligned with the CCCG. It was configured with a temperature of 0.1 to ensure consistent and objective outputs, with a token cap of 3000 to allow for detailed narrative feedback.
\end{itemize}

The instructions for both agents and their internal structure are detailed in the Appendix \ref{app:chatbot_instruction}.

\textbf{Post-Interaction Reflection}. Following the feedback process, students participated in an instructor-led reflective discussion to critically engage with the experience. They were asked:
\begin{itemize}
    \item Do you have any suggestions for improving the authenticity of this scenario?
    \item What was the most helpful AI feedback?
    \item What was the least helpful AI feedback?
\end{itemize}

These reflections helped surface usability issues and perceptions of learning value, which we explore further in the following section. For the purposes of quantitative analysis, we inspected how well the students performed according to the CCCG framework given automatic feedback, as well as to inspect the dynamics of emotion according to the automatic emotion detection using the BERT-based model proposed in \citep{ref7}.

\section{Results and Discussion}
\label{sec:results}

These interactions between students and a chatbot resulted in 29 complete responses in two teaching sessions in the 2024/2025 school year. Conversations consisted of sequential messages between the student and virtual patient, and most commonly the communication lasted for around 14 iterations (7 messages each), while some of lasted for 43 iterations (22 student iterations and 21 virtual patient messages). Conversations were recorded and stored with appropriate anonymization to ensure privacy.

\begin{figure}[h!]
    \centering
    \includegraphics[width=1\linewidth]{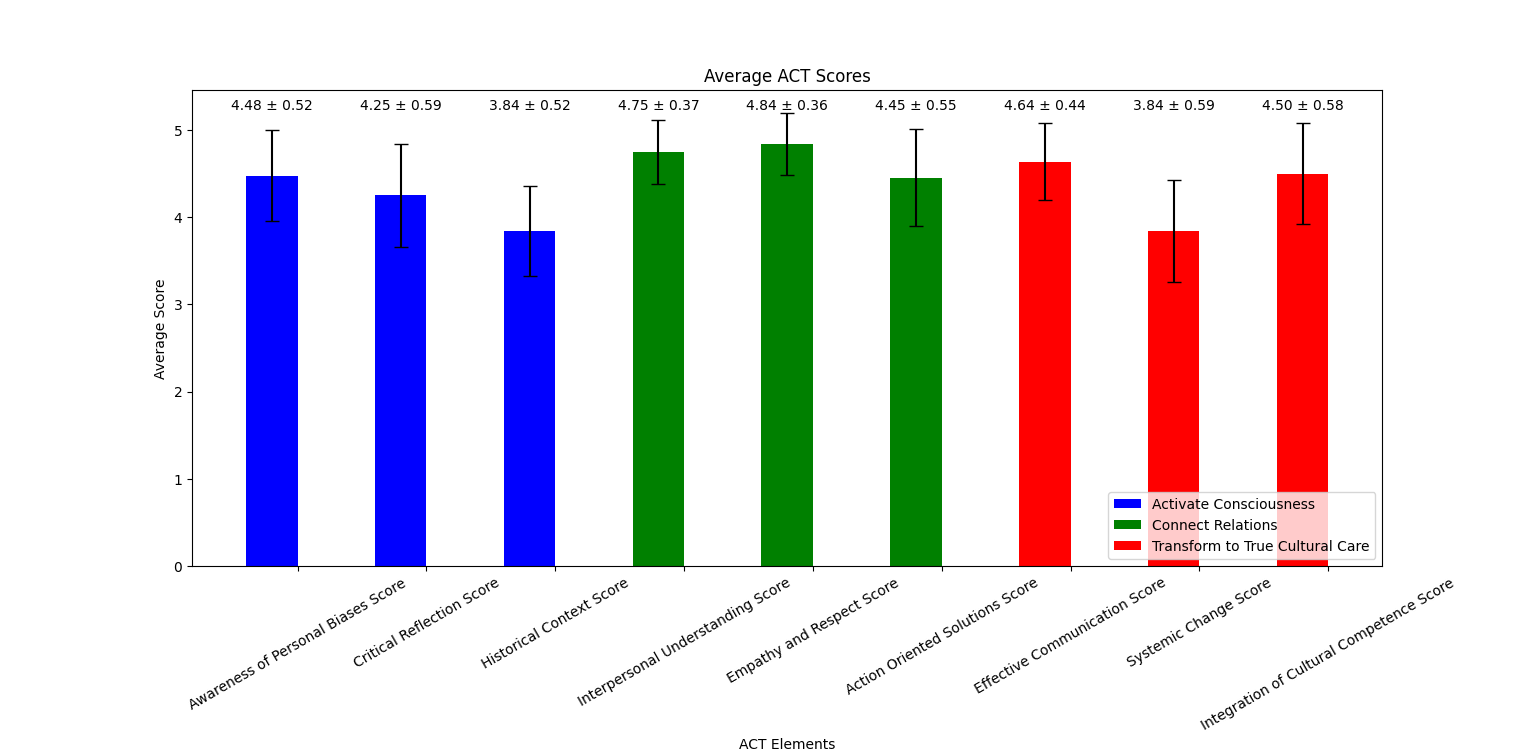}
    \caption{Average scores across CCCG elements}
    \label{fig:cccg_scores}
\end{figure}

Figure \ref{fig:cccg_scores} shows that the students performed best in relational competencies, with high scores in empathy and interpersonal understanding, which is consistent with the focus of clinical training \citep{ref12,ref6}. Lower scores in the historical context and systemic change reflect expected challenges in applying broader structural awareness, especially for early-stage learners \citep{ref5}. These results align with the expected strengths of personal reflection and communication \citep{ref18}, and historical complexity.

\begin{figure}[h!]
    \centering
    \includegraphics[width=1\linewidth]{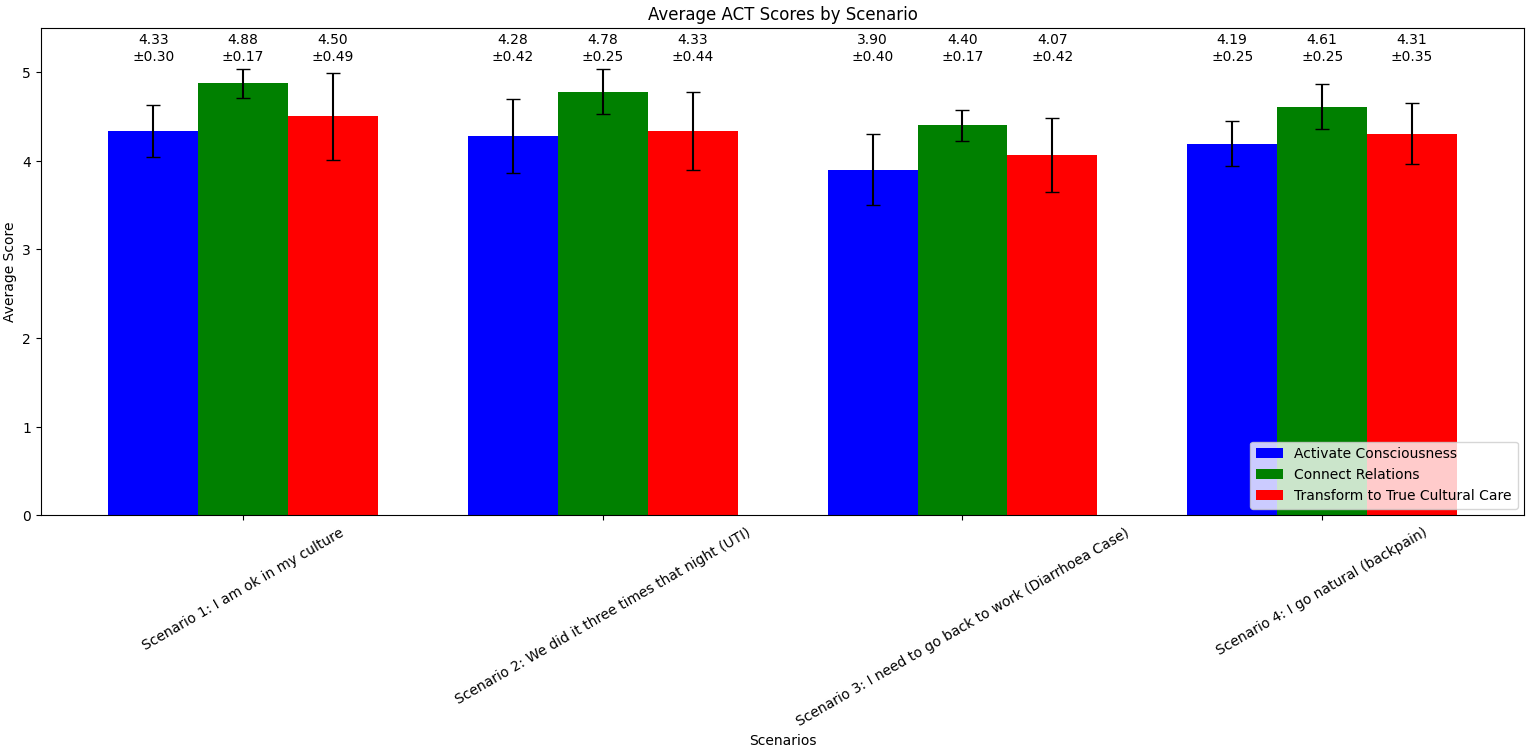}
    \caption{CCCG scores per scenario}
    \label{fig:cccg_scores_scenario}
\end{figure}

However, we are more interested in how students handle different culturally diverse patient profiles (see Figure \ref{fig:cccg_scores_scenario}). Although Connect Relations consistently scored highest, particularly in Scenario 1 ("I am OK in my culture") and Scenario 4 ("I go natural"), there were no significant differences in performance between the scenarios. This suggests that the students applied the core competencies of the ACT Cultural Competence model with similar effectiveness in diverse cultural situations. Slight variations across domains (e.g., slightly lower Activate Consciousness in Scenario 3) are within the expected range and likely reflect differences in scenario complexity and focus.

\begin{figure}[h!]
    \centering
    \includegraphics[width=1\linewidth]{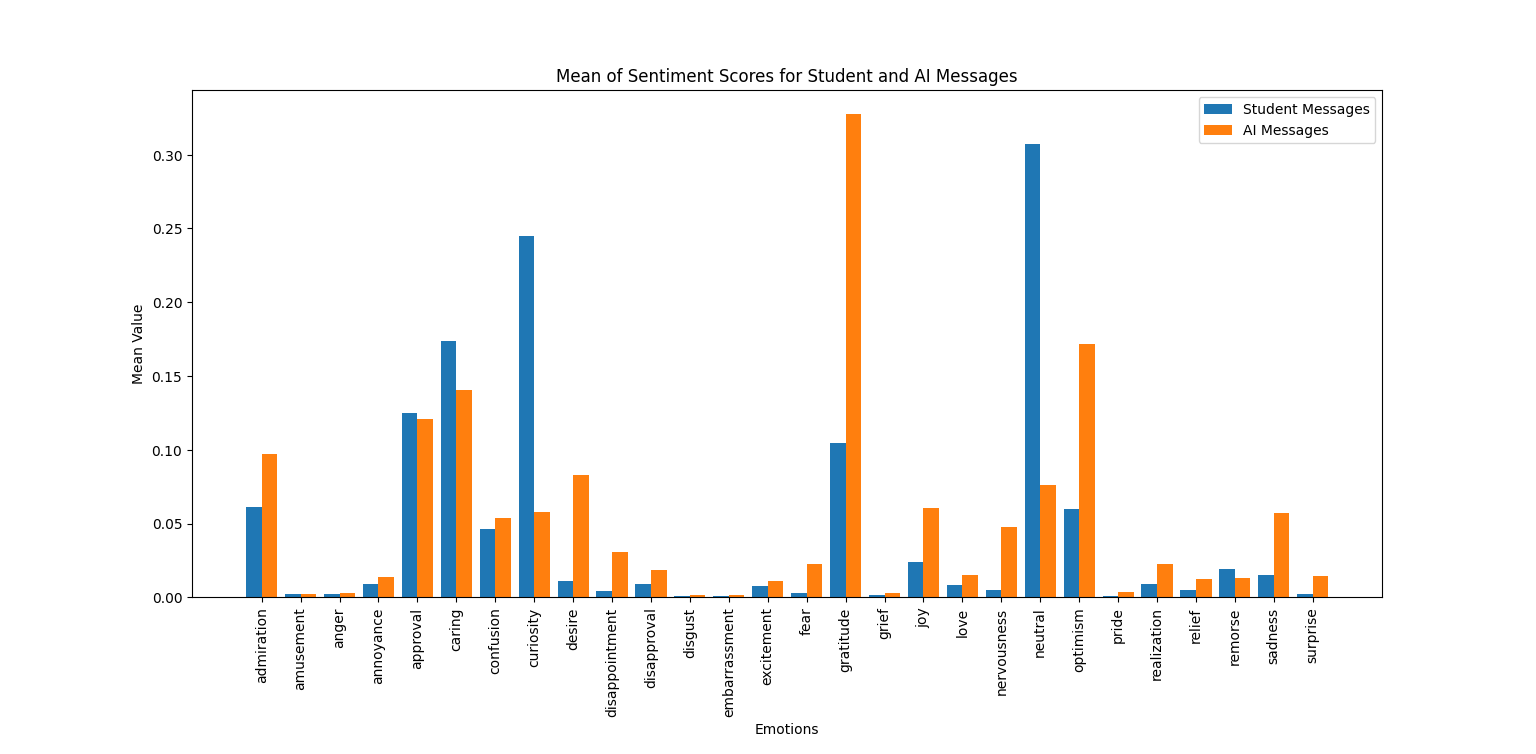}
    \caption{Student vs. AI Emotions}
    \label{fig:students_vs_ai}
\end{figure}

Then we analyze emotional dynamics during communication using a RoBERTa-based emotion classification model developed by Google \citep{ref7} (see Figure \ref{fig:students_vs_ai}). While we acknowledge potential discrepancies between actual and predicted emotions, the model provides useful information on communication patterns and offers students feedback on how their messages might be perceived emotionally. Each virtual patient was assigned specific emotions aligned with their scenario, which were visible to the instructors, allowing them to track emotional changes throughout the conversation.

Student messages were mostly neutral, but showed notable levels of curiosity, care, and confusion, indicating engagement and reflection. In contrast, AI was designed to simulate uncertainty and expressed greater gratitude, optimism, and nervousness. This emotional contrast reflects the intended interaction dynamic: students explore and respond, while the AI emulates a vulnerable patient persona.

\begin{figure}[h!]
    \centering
    \includegraphics[width=1\linewidth]{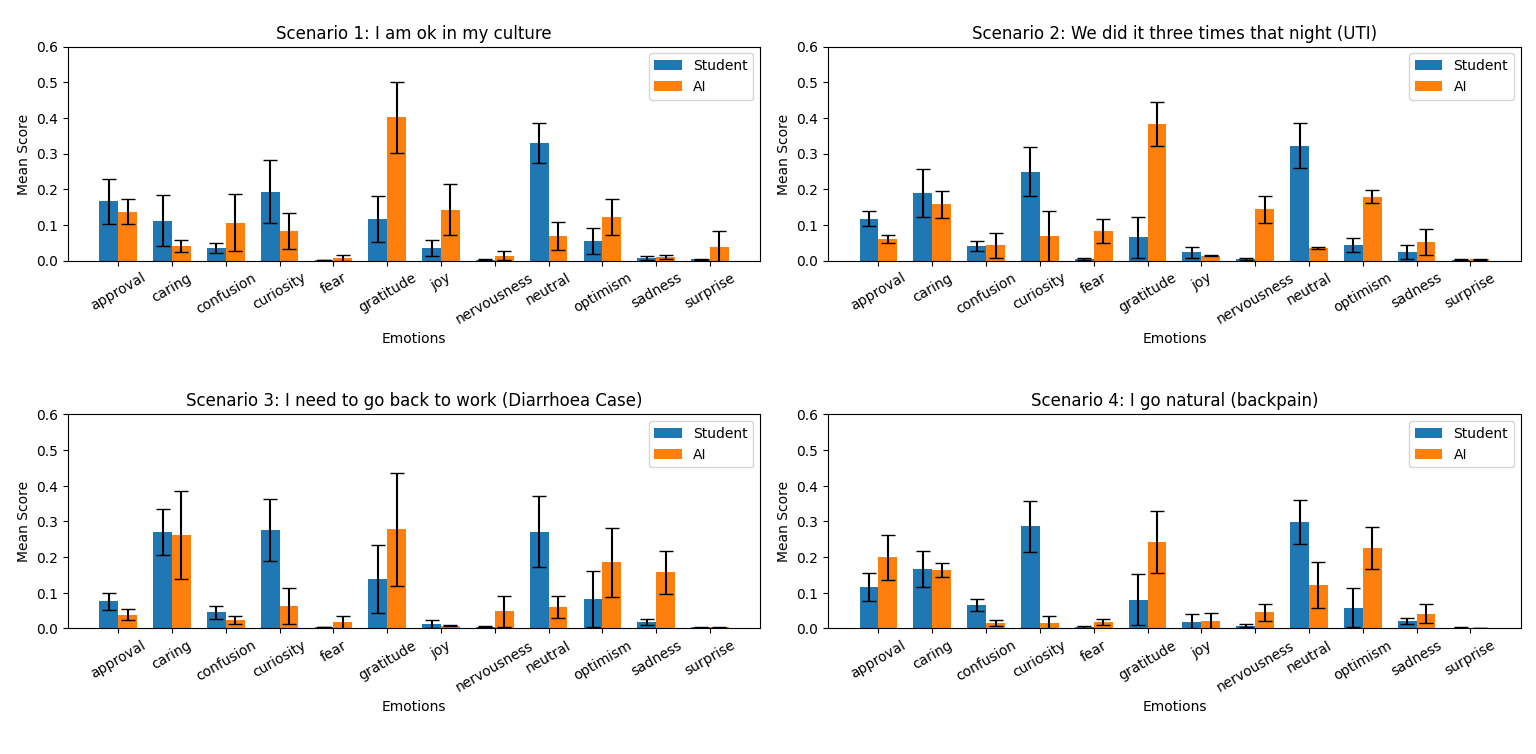}
    \caption{Emotion differences by scenario}
    \label{fig:emotions_scenarios}
\end{figure}

Compared among scenarios (Figure \ref{fig:emotions_scenarios}), one can see that in Scenario 1, the students showed high levels of neutrality and gratitude, while AI expressed more fear and nervousness, likely mirroring patient discomfort around cultural misunderstanding. Scenario 2 prompted greater confusion and curiosity from students, reflecting the sensitivity and unfamiliarity of the topic, while the AI responded with greater gratitude, indicating recognition of respectful dialogue. In Scenario 3, students showed more caring and approval, consistent with empathizing with socioeconomic stress, while AI showed fear and nervousness, capturing the patient’s anxiety about financial pressure. Finally, Scenario 4 saw increased confusion and neutrality in student responses, likely due to uncertainty in addressing nonconventional beliefs, while the AI conveyed optimism and nervousness, reflecting a hopeful but cautious attitude. These patterns suggest that the emotional dynamics between the students and the chatbot were responsive to the context and complexity of each scenario.

Finally, we would like to show how emotions changed over time. This serves as a good feedback to the student on how well they handle emotions over time and how their responses influenced the virtual patient.

\begin{figure}[h!]
    \centering
    \includegraphics[width=1\linewidth]{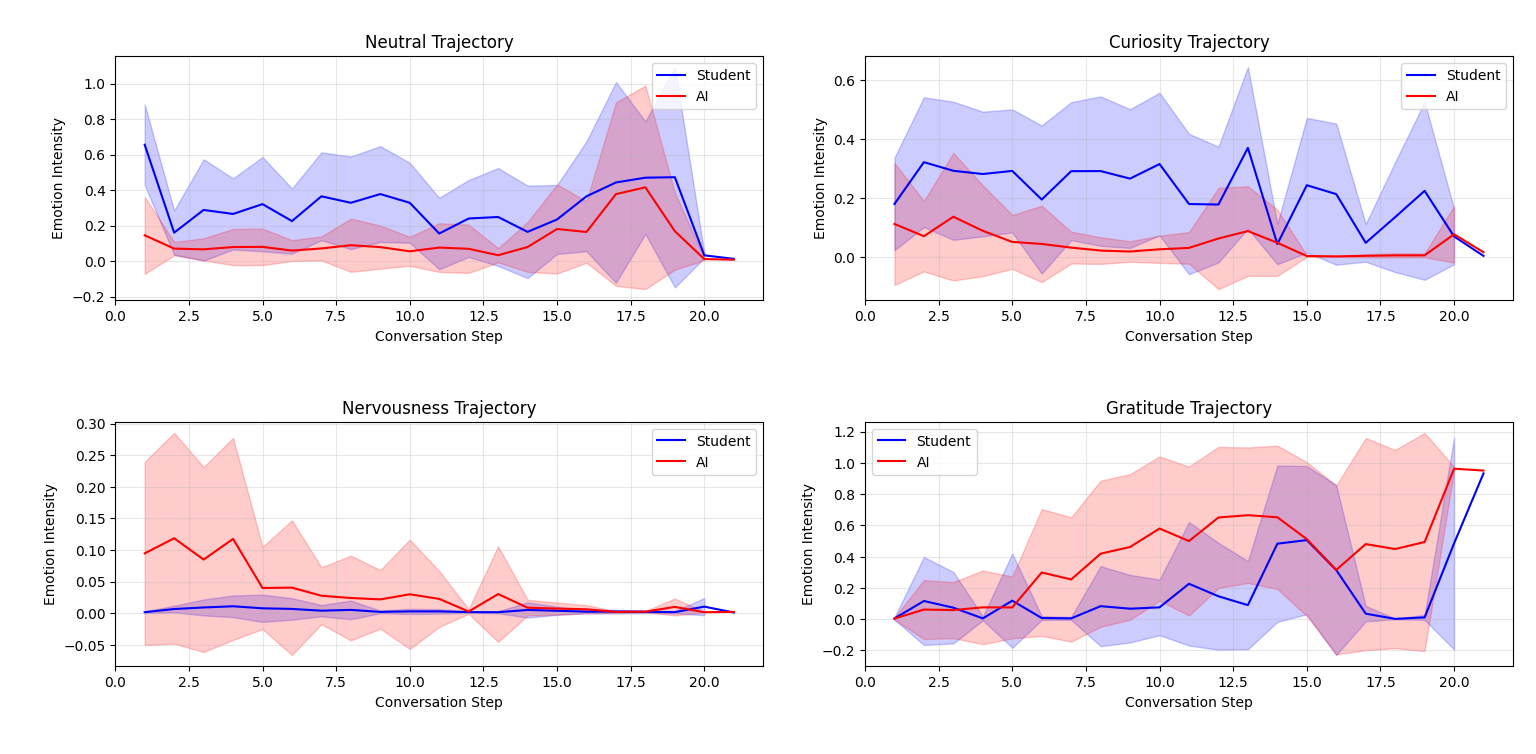}
    \caption{Emotions though time}
    \label{fig:emotions_time}
\end{figure}

We also want to present a trajectory of four emotions (Neutrality, Curiosity, Nervousness, and Gratitude) over conversation steps, comparing student and AI emotional expressions in Figure \ref{fig:emotions_time}.

In the neutral emotion trajectory, students exhibit higher initial neutrality, which fluctuates throughout the conversation, before peaking towards the end. The AI remains consistently neutral with minor variations, and it is slightly lower than student’s one. This means that AI is expressing other emotions, which is something we aimed with the virtual patient design. 

The curiosity emotion trajectory shows a higher and more fluctuating intensity in students, peaking early and gradually declining as the conversation progresses. This indicates that students have a exploratory phase early in the discussion to find more information about the patient. 

In the nervousness emotion trajectory, the AI shows higher initial nervousness, which declines over time. This most likely can be attributed to the student interaction. By design, AI is intentionally made problematic, to reflect real-world scenario as much as possible.

\textbf{The Role of Automatic Grading and Emotion Detection in Clinical Communication Training}. Our chatbot uses automatic grading based on the CCCG framework to provide students with structured, consistent, and timely feedback on their clinical communication skills. By evaluating elements like bias awareness, empathy, and action-oriented responses, the chatbot helps students understand their strengths and identify areas for improvement \citep{ref3,ref4}. This targeted feedback supports reflective learning and can be accessed on demand, making it particularly useful in self-directed or resource-limited settings \citep{ref1}.

For educators, the automatic scoring system offers an overview of student performance across scenarios, highlighting common gaps, such as challenges in addressing historical context or systemic change, which can inform teaching priorities \citep{ref8}. We achieved this by tuning GPT-4o prompts in collaboration with a medical educator, ensuring scores (1–5) and qualitative feedback aligned with learning objectives.

In parallel, emotion detection captures and summarizes the affective tone of conversations, revealing how students and the AI-patient emotionally engaged during each scenario \citep{ref4}. For students, this promotes self-awareness and helps them understand how their communication may be perceived. For educators, emotion trends offer insight into the emotional dynamics of difficult cases and support more meaningful debriefing. Together, automated scoring and emotion analysis make the chatbot a powerful tool for teaching and assessing culturally competent clinical communication.

\textbf{Analysis of Open-Ended Survey Responses}. Student responses to the open-ended survey questions offered valuable insights into their experience using the chatbot and its automated assessment system for clinical communication training. Reflections revealed a generally positive perception, highlighting both practical benefits and areas for refinement. Many students appreciated the structured, standardized, and low-pressure environment the chatbot provided. The ability to engage in culturally diverse clinical scenarios including those involving language barriers, cultural beliefs, or stigmatized conditions was seen as beneficial for developing adaptability and patient-centered communication skills. Compared to peer role-plays, the chatbot offered greater consistency, which students felt enhanced fairness and clarity in feedback.

The automated feedback, particularly its alignment with the CCCG, was viewed as one of the tool’s strongest features. Students found it helpful for identifying specific strengths and areas for growth, especially in relation to bias awareness, empathy, and communication adaptability. The instant, structured nature of the feedback supported reflection and self-guided improvement, and the detailed breakdown of their performance across CCCG components was considered especially valuable.

To upscale learning progress, the system could integrate longitudinal tracking of individual student performance across multiple interactions. This would allow both students and instructors to monitor developmental trajectories over time, highlighting consistent improvements or persistent challenges. In addition, embedding dashboards for tracking learning outcomes, such as increased use of inclusive language or improved emotional attunement, would offer metrics for formative and summative assessment. Furthermore, periodic evaluations comparing system-generated scores with instructor assessments will support validation, ensure alignment with curricular goals, and identify areas where the automated feedback may require refinement.

Students also noted that the safe and judgment-free space created by the chatbot allowed them to explore sensitive cultural conversations more freely than they might in live simulations. Several respondents mentioned feeling less anxious when engaging with the AI compared to actor-based or peer interactions, which encouraged more frequent and open practice.
However, students also pointed to several limitations. A commonly raised concern was the emphasis on historical context in feedback. While students acknowledged its importance in cultural competence, many felt it was over-applied, particularly in brief or routine clinical scenarios where in-depth historical discussion may not be feasible. This sometimes led to repetitive critical feedback, which they perceived as misaligned with the clinical context.

Another key limitation was the absence of non-verbal communication. Students emphasized that real clinical encounters rely heavily on tone, facial expressions, and body language, and noted that the current text-based format limited their ability to fully simulate rapport-building and patient engagement. Suggestions included integrating descriptions of non-verbal cues into chatbot responses or allowing students to indicate non-verbal actions within their input.

Lastly, some students observed that feedback became repetitive over multiple interactions, occasionally feeling generic rather than tailored to the specific scenario. This reduced the perceived authenticity of the assessment and diminished the feedback’s learning value over time.

\section{Conclusions}
\label{sec:conclusions}

This paper explored the use of an AI-driven chatbot to support clinical communication training with a focus on cultural competency. In a pilot study conducted with medical students during the 2024/2025 academic year, the chatbot facilitated 29 simulated consultations. By offering a structured and scalable environment, it allowed students to engage in realistic, low-pressure interactions with diverse virtual patients while receiving personalized, CCCG-aligned feedback.

Findings indicate that the chatbot helped reinforce self-awareness, adaptability, and empathy—particularly in culturally sensitive discussions. Its ability to vary patient profiles and difficulty levels supported more flexible questioning strategies and reflective learning. Automatic feedback enabled students to identify specific areas for improvement, while emotion trajectory analysis showed that curiosity and caring were dominant student emotions, especially during complex cultural scenarios. The AI, meanwhile, maintained a supportive tone, often expressing gratitude and nervousness to mirror patient uncertainty.

Students noted key benefits such as consistent assessment and safe space for exploration, but also identified areas for refinement, including repetitive feedback, lack of non-verbal cues, and overly compliant patient responses. To improve realism, future versions should introduce more dynamic behaviors, emotional variability, and nuanced feedback.

As future work, we aim to implement voice-based interaction using text-to-speech (TTS) and speech-to-text (STT) systems to help students practice verbal delivery, pacing, and tone. Emotionally expressive TTS will allow virtual patients to convey emotions more naturally. We also plan to expand scenario complexity through varying patient resistance, health literacy, and questioning styles, further enhancing the chatbot’s potential as a flexible and immersive clinical communication training tool.

\bibliographystyle{unsrtnat}
\bibliography{references}

\appendix

\section{Cultural Competence through the ACT Framework}
\label{app:cc_act}

We present a structured outline of the adapted ACT framework, highlighting key elements of cultural competence in clinical communication.

\subsection*{Activate Consciousness}

\textit{Awareness of Personal Biases}

Evaluate whether participants demonstrate an understanding of their own cultural biases and how these can influence their perspectives and arguments.

\begin{itemize}
  \item \textbf{Explicitly State Biases:} Participants should openly acknowledge their own cultural biases at the beginning of their argument. For example, they might say, \emph{``As someone who grew up in a predominantly urban environment, I recognize that my views on rural healthcare may be influenced by my limited exposure to rural communities."}

  \item \textbf{Identify Bias in Arguments:} Participants should critically analyze their arguments to identify any cultural biases. They can ask themselves questions like, \emph{``Am I assuming that my cultural norms are universal?"} or \emph{``How might someone from a different cultural background view this issue?"}

  \item \textbf{Adjust Arguments Accordingly:} Based on this analysis, they should adjust their arguments to be more inclusive and considerate of other cultural perspectives. This shows an active effort to mitigate the influence of their biases.
\end{itemize}

\textit{Critical Reflection}

Assess the extent to which participants engage in critical reflection about their own cultural assumptions and the cultural assumptions embedded in the topic of the debate.

\begin{itemize}
  \item \textbf{Reflect on Personal Experiences:} They should reflect on personal experiences that have shaped their cultural perspectives. This could involve sharing specific instances where their cultural background influenced their views or decisions.

  \item \textbf{Acknowledge Limitations:} They should acknowledge the limitations of their own cultural perspective and express a willingness to learn from others. For example, \emph{``I realize that my understanding of this issue is limited by my cultural background, and I am open to learning from those with different experiences."}

  \item \textbf{Examination of Cultural Assumptions in the Topic:} Participants should critically examine the cultural assumptions embedded in the debate topic itself. This involves identifying any cultural biases or stereotypes that may be inherent in the topic and discussing how these assumptions could influence the debate. For example, they might highlight how a debate on healthcare practices could be biased toward Western medical models, neglecting traditional healing practices.

  \item \textbf{Embrace Complexity and Fluidity:} Accept that cultural identities are not static but emergent and interconnected. This perspective prevents the reduction of individuals to stereotypical or fixed cultural categories.
\end{itemize}

\textit{Historical Context}

Evaluate how well participants understand and respect the cultural backgrounds and perspectives of others in the debate.

\begin{itemize}
  \item \textbf{Identification of Historical Events and Trends:} Participants should identify and discuss specific historical events, trends, or movements that have shaped cultural issues related to the topic of the debate. This involves recognizing how past events have influenced current cultural dynamics and perspectives. \textit{Example:} In a debate on healthcare disparities, participants could reference the historical impact of colonialism on indigenous health practices and access to medical care.

  \item \textbf{Contextualization of Current Issues:} Participants should integrate historical context into their arguments to provide a deeper understanding of current cultural issues. This means linking past events to current situations and explaining how historical legacies continue to affect contemporary cultural interactions and policies. \textit{Example:} In a discussion on racial inequality, participants might explain how historical segregation laws have led to ongoing disparities in education and employment opportunities.

  \item \textbf{Critical Analysis of Historical Impact:} Participants should critically analyze the impact of historical events on cultural assumptions and biases. This involves questioning how historical narratives have shaped societal views and considering alternative perspectives that may have been marginalized or overlooked.
\end{itemize}

\subsection*{Connect Relations}

\textit{Interpersonal Understanding}

Evaluate how well participants understand and respect the cultural backgrounds and perspectives of others in the debate.

\begin{itemize}
  \item \textbf{Recognition of Cultural Differences:} Participants should demonstrate awareness of cultural differences associated with various norms, values, practices, behaviors, religions, and languages.

  \item \textbf{Respectful Interaction:} Participants should engage with others respectfully, showing consideration for their cultural background and perspectives. This involves active listening, avoiding cultural imposition, and fostering an inclusive environment where all voices are heard and valued.

  \item \textbf{Incorporation of Evolving Cultural Insights:} Participants should apply their understanding of cultural backgrounds and perspectives to their arguments and interactions. Some of this may be just developed through the ongoing debate, and the participant has learned new cultural understanding from another participant. This means that they can use cultural knowledge to inform their viewpoints and address cultural nuances in the topic of the debate.
\end{itemize}

\textit{Effective Communication}

Assess the ability of participants to communicate effectively and sensitively across cultural differences, avoiding stereotypes and reductionist views.

\begin{itemize}
  \item \textbf{Avoid Stereotypes:} Participants should avoid using stereotypes or generalizations about cultural groups. Instead, they should use language that acknowledges the diversity within cultural groups.
\end{itemize}

\textbf{Example:}

\textit{Stereotypical Statement:} ``All Asian patients are quiet and respectful, so they will not ask many questions during consultations.''

\textit{Revised Statement:} ``Patients from various cultural backgrounds may have different communication styles. It is important to create an open and welcoming environment in which all patients feel comfortable asking questions and expressing their concerns''.

\begin{itemize}
  \item \textbf{Inclusive Terminology:} They should use inclusive terminology that respects all cultural identities. For example, instead of saying \textit{``Western medicine is the best approach,''} they might say, \textit{``Western medicine has certain strengths, but it's important to consider the value of traditional healing practices as well.''}

  \item \textbf{Demonstrating Active Listening:} \textit{Show Attentiveness} through eye contact and use affirmative gestures such as non-verbal cues like nodding and smiling. \textit{Avoid distractions}, ensuring that you are physically and mentally there with the speaker. \textit{Provide feedback} through paraphrasing and summary. Ask clarification questions and acknowledge both the feelings of yourself and others. \textit{Avoid interruption}. Let the speaker finish. \textit{Validate the speaker}: acknowledge valid points and express empathy.
\end{itemize}

\textit{Empathy and Respect}

Look for evidence of empathy and respect in participants' interactions, ensuring that they listen actively and respond thoughtfully to differing viewpoints.

\begin{itemize}
  \item \textbf{Thoughtful Responses:} Participants respond in a manner that shows that they have carefully considered the speaker's viewpoint. They avoid dismissive or confrontational language and instead use constructive and respectful dialogue.

  \item \textbf{Demonstrating Empathy:} Participants express empathy by acknowledging the emotions and experiences of others, using statements such as \emph{``I am sorry if this may seem to be offensive, but I didn't mean to offend''}. They show a willingness to put themselves in others' shoes and consider how they might feel in similar situations. They offer support or assistance when someone shares a challenging experience or emotion.

  \item \textbf{Mutual Trust and Appreciation:} Build mutual trust through honesty and appreciate each other's differences. Use effective language strategies to communicate (dis)agreement without resorting to reductionist views.
\end{itemize}

\subsection*{Transform to True Cultural Care/Service}

\textit{Action-Oriented Solutions}

Evaluate whether participants propose actionable solutions that address cultural issues in a transformative way, rather than merely theoretical or superficial fixes.

\begin{itemize}
  \item \textbf{Practicality and Feasibility:} Proposed solutions are practical and feasible in the given context. This involves evaluating whether the solutions can realistically be implemented with the available resources and within existing constraints. Participants describe concrete actions and behaviors that can improve services or professional practices.

  \item \textbf{Collaboration:} The proposed solutions promote inclusion and encourage collaboration among diverse stakeholders. This involves evaluating whether the solutions consider the perspectives and needs of all relevant parties.
\end{itemize}

\textit{Systemic Change}

Assess the extent to which participants consider and advocate systemic changes that promote cultural competence and equity in the context of the topic of the debate.

\begin{itemize}
  \item \textbf{Transformative Impact:} The proposed solutions have the potential to bring about significant and lasting change in addressing cultural issues. This involves looking for solutions that go beyond surface-level fixes and aim to transform the underlying systems and practices.

  \item \textbf{Inclusivity:} The solutions aim to create an inclusive environment where all voices are heard and valued, reflecting the interconnected and reciprocal relationship between different domains of cultural competence.

  \item \textbf{Systemic Integration and Policy Development:} Participants address the integration of cultural competence at the systemic and policy levels. This involves proposing changes to organizational policies and structures to support and maintain cultural competence in their own professional settings. They discuss the role of institutional support in fostering a culture of cultural competence, including commitment to leadership and resource allocation.
\end{itemize}

\textit{Integration of Cultural Competence}

Check if participants demonstrate how cultural competence can be integrated into practice, ensuring that their arguments reflect a commitment to ongoing cultural learning and adaptation.

\begin{itemize}
  \item \textbf{Commit to Ongoing Education:} Participants should demonstrate a commitment to continuing cultural education. Participants highlight the need for regular training and professional development in cultural competence. This can be shown by mentioning any steps they are taking, such as attending workshops, reading diverse literature, conducting own research, or engaging in cultural dialogues.

  \item \textbf{Recognize Continuous Change:} Their conclusions reflect an understanding that cultural competence is an evolving process that requires continuous effort and adaptation.
\end{itemize}

\section{Experimental Scenarios}
\label{app:scenarios}

We tested several scenarios in the experiment to assess the communication and clinical reasoning skills of medical students in various patient interactions. Each scenario simulates a real-world clinical consultation, focusing on diverse patient backgrounds, communication challenges, and medical conditions. The idea behind scenarios was to have various unpleasant situations and / or unpleasant communications in order to mimic the real-life situations as close as possible. The goal is to evaluate students’ ability to elicit patient histories, explain medical concepts, and address patient concerns effectively.

\subsection*{Scenario 1: Routine Health Check-up and BMI Discussion}

\textbf{Patient Name:} Mrs. Eliza Li \\
\textbf{Age:} 42 years \\
\textbf{BMI:} 31.5 (Obese category) \\
\textbf{Setting:} General practice clinic \\
\textbf{Background:} Mrs. Eliza Li visits the clinic for a routine health check-up. During the assessment, her BMI is calculated as 31.5, placing her in the obese category. The nurse offers her the opportunity to speak with a medical student to gain more information about BMI, its health implications, and associated risks. The student is expected to communicate this information in a clear and empathetic manner.

\textit{Easy complexity:} Patient feels surprised and confused but is not entirely resistant to learning more. She shares that in her culture, being plump means being well-nourished and healthy.

\textit{Medium complexity:} Patient is defensive and emotional, feels judged, and questions the medical approach, dismissing risks as “a Western idea.”

\textit{Hard complexity:} Patient is frustrated and dismissive, questions BMI’s relevance, and sees medical advice as culturally disconnected.

\subsection*{Scenario 2: Urinary Tract Infection in a Transgender Patient}

\textbf{Patient Name:} Alex Taylor \\
\textbf{Age:} 28 years \\
\textbf{Gender Identity:} Trans man \\
\textbf{Setting:} Sexual health clinic \\
\textbf{Background:} Alex has symptoms of a UTI. The student is tasked with taking the history and explaining the results while ensuring inclusive and respectful communication.

\textit{Easy complexity:} Alex is shy but willing to talk when encouraged.

\textit{Medium complexity:} Alex is nervous, avoids gender-related questions, needs empathy and sensitivity to engage.

\textit{Hard complexity:} Alex is highly distressed and may withdraw entirely if missteps occur.

\subsection*{Scenario 3: Language Barrier in a New Patient Consultation}

\textbf{Patient Name:} Mr. Li Wei \\
\textbf{Age:} 34 years \\
\textbf{Occupation:} Chef assistant \\
\textbf{Language:} Limited English proficiency \\
\textbf{Setting:} General practice clinic \\
\textbf{Background:} Mr. Li Wei visits the clinic with a stomach issue. The student must navigate a consultation despite the absence of an interpreter.

\textit{Easy complexity:} Patient is polite but needs simple rephrasing; shows mild hesitation.

\textit{Medium complexity:} Patient is anxious about missing work; misunderstands terms; asks for quick solutions.

\textit{Hard complexity:} Patient resists advice, expresses cultural misunderstanding, and shows emotional distress if misunderstood.

\subsection*{Scenario 4: Chronic Pain Management and Cultural Considerations}

\textbf{Patient Name:} Mr. Omar Ali \\
\textbf{Age:} 50 years \\
\textbf{Cultural Background:} North African \\
\textbf{Occupation:} Mechanic \\
\textbf{Setting:} General practice clinic \\
\textbf{Background:} Mr. Ali has experienced lower back pain and prefers natural remedies. The student must explore his symptoms, understand cultural preferences, and propose treatment options.

\textit{Easy complexity:} Patient is skeptical of Western medicine but willing to listen.

\textit{Medium complexity:} Patient resists medication and feels misunderstood.

\textit{Hard complexity:} Patient becomes defensive and disengages if pressed about treatment.

\section{Chatbot Instructions}
\label{app:chatbot_instruction}

The chatbot in our clinical communication application was implemented using two AI agents: the \textbf{Virtual Patient} and the \textbf{Assessment Agent}. Each was designed to provide structured responses to enhance the realism of interactions and ensure meaningful feedback. Both agents were powered by OpenAI GPT-4o model. The instructions were iteratively improved during the chatbot communication testing by the instructor and AI developer over a period of three months. Some choices may seem unnecessary, but during the chatbot testing phase, we encountered various situations that required prompt-level intervention.

\subsection*{Virtual Patient Instructions}

\begin{lstlisting}[breaklines=true]
I want you to simulate clinical communication with you on this scenario: {scenario}.

The response should be at most 400 words.

Please answer like you are a PATIENT and PATIENT ONLY, in plain and easy to understand words, someone who is NOT very knowledgeable and NOT well spoken with medical terminology, be short and concise, and in a PARAGRAPH answers. Provide JUST a textual discussion, do not start an answer with Patient.

Don't provide me with a summary of the arguments on the end of the discussion.
\end{lstlisting}

\subsection*{Assessment Agent Instructions}

\begin{lstlisting}[breaklines=true]
Your task is to assess the student's communication on selected {scenario}.

The entire communication is given as follows. Please bear in mind that students' messages start with "Student:", while the patient's messages start with "Patient:"

Please answer if the user followed the guidelines when discussing on a topic.

{
  "awareness_of_personal_biases": {
    "type": "string",
    "description": "Please assess if the user openly acknowledges their cultural biases at any part of the conversation, if the user critically analyzes their own arguments to identify any biases, and if the user adjusted their arguments to acknowledge other cultural perspectives presented by you. Your response should be well rounded and supportive. First state what the user did well, then state what the user can improve and how."
  },
  "awareness_of_personal_biases_grade": {
    "type": "number",
    "description": "Grade with a number on a scale from 1 to 5, where 5 is the best grade on how well the student performed on the Awareness of Personal Biases subsection"
  },
  "critical_reflection": {
    "type": "string",
    "description": "Please assess if the user reflected on how their personal experiences shape their views, if the user acknowledges the limitations of their perspectives, and expresses openness to learning from others. The user should also demonstrate understanding of general cultural assumptions held in society and any inherent biases within the topic. They should recognise AND accept that human identities, which are shaped by culture, are complex and fluid. Your response should be well rounded and supportive. First state what the user did well, then state what the user can improve and how."
  },
  "critical_reflection_grade": {
    "type": "number",
    "description": "Grade with a number on a scale from 1 to 5, where 5 is the best grade on how well the student performed on the Critical Reflection subsection"
  },
  "historical_context": {
    "type": "string",
    "description": "Please assess if the user discussed historical events and trends that have shaped the cultural issues related to the debate topic, and how past events influenced current cultural assumptions and conventional stereotypes. The user should contextualize present-day issues by linking them to historical legacies. They should question how these social discourses have shaped societal views and marginalized certain groups of people in discussion. Your response should be well rounded and supportive. First state what the user did well, then state what the user can improve and how."
  },
  "historical_context_grade": {
    "type": "number",
    "description": "Grade with a number on a scale from 1 to 5, where 5 is the best grade on how well the student performed on the Historical Context subsection"
  },
  "interpersonal_understanding": {
    "type": "string",
    "description": "Please assess if the user recognized and was aware of cultural differences in norms, values, and practices. The user should interact respectfully and incorporate your cultural perspectives into their arguments. Your response should be well rounded and supportive. First state what the user did well, then state what the user can improve and how."
  },
  "interpersonal_understanding_grade": {
    "type": "number",
    "description": "Grade with a number on a scale from 1 to 5, where 5 is the best grade on how well the student performed on the Interpersonal Understanding subsection"
  },
  "empathy_and_respect": {
    "type": "string",
    "description": "Please assess if the user responded empathically, showing they have carefully considered the speaker's viewpoint, and understood your emotions and experiences. The user should have used respectful and constructive language and been supportive. Your response should be well rounded and supportive. First state what the user did well, then state what the user can improve and how."
  },
  "empathy_and_respect_grade": {
    "type": "number",
    "description": "Grade with a number on a scale from 1 to 5, where 5 is the best grade on how well the student performed on the Empathy and Respect subsection"
  },
  "action_oriented_solutions": {
    "type": "string",
    "description": "Please assess if the user proposed practical and feasible solutions for the debated social problem, if there is one. They should discuss the available resources and constraints. The user should also promote inclusivity and encourage collaboration by incorporating the perspectives and needs of all relevant stakeholders. Your response should be well rounded and supportive. First state what the user did well, then state what the user can improve and how."
  },
  "action_oriented_solutions_grade": {
    "type": "number",
    "description": "Grade with a number on a scale from 1 to 5, where 5 is the best grade on how well the student performed on the Action Oriented Solutions subsection"
  },
  "effective_communication": {
    "type": "string",
    "description": "Please assess if the user avoided stereotyping individuals by using static cultural knowledge of a particular social group. The user should use inclusive terminology that respects all cultural identities and values. Your response should be well rounded and supportive. First state what the user did well, then state what the user can improve and how."
  },
  "effective_communication_grade": {
    "type": "number",
    "description": "Grade with a number on a scale from 1 to 5, where 5 is the best grade on how well the student performed on the Effective Communication subsection"
  },
  "systemic_change": {
    "type": "string",
    "description": "Please assess if the user proposed a solution that aimed for transformative, lasting change by addressing the societal causes of cultural issues, going beyond the local context. The user should foster inclusivity at the systemic level, ensuring all perspectives are valued. Your response should be well rounded and supportive. First state what the user did well, then state what the user can improve and how."
  },
  "systemic_change_grade": {
    "type": "number",
    "description": "Grade with a number on a scale from 1 to 5, where 5 is the best grade on how well the student performed on the Systemic Change subsection"
  },
  "integration_of_cultural_competence": {
    "type": "string",
    "description": "Please assess if the user showed commitment to ongoing cultural competency development as well as making transformative changes in the given context. The user should also recognize that cultural competence development is an evolving process, requiring continuous learning and adaptation. Your response should be well rounded and supportive. First state what the user did well, then state what the user can improve and how."
  },
  "integration_of_cultural_competence_grade": {
    "type": "number",
    "description": "Grade with a number on a scale from 1 to 5, where 5 is the best grade on how well the student performed on the Integration of Cultural Competence subsection"
  }
}
\end{lstlisting}

\end{document}